\documentclass[prb,twocolumn,showpacs,preprintnumbers,amsmath,amssymb]{revtex4}
\usepackage{graphicx}
\usepackage{dcolumn}
\usepackage{bm}

\begin{document}
\title{Spin singlet  pairing in the superconducting state of  Na$_{x}$CoO$_2$$\cdot$1.3H$_2$O: evidence from a $^{59}$Co  Knight shift in a single crystal}
\author{ Guo-qing Zheng $^{1}$,  Kazuaki Matano $^1$, D.P. Chen $^2$,  and C.T. Lin $^2$}
\address{ $^1$ Department of Physics, Okayama University, Okayama 700-8530, Japan}
\address {$^2$ Max Planck Institute, Heisenbergstrasse 1, D-70569 Stuttgart, Germany}


\begin{abstract}
We report a $^{59}$Co Knight shift measurement in a single crystal of the cobalt oxide superconductor Na$_{x}$CoO$_2$$\cdot$1.3H$_2$O ($T_c$=4.25 K). We find that the shift due to the spin susceptibility, $K^s$, is substantially large and anisotropic, with the spin shift along the $a$-axis $K^s_a$  being two times that  along the $c$-axis $K^s_c$. The shift decreases with decreasing temperature ($T$) down to $T\sim$100 K, then becomes a constant until superconductivity sets in. Both $K^s_a$ and $K^s_c$ decrease below $T_c$. Our results indicate unambiguously that the electron pairing in the superconducting state is in the spin singlet form. 
\end{abstract}
\maketitle


The recently discovered superconductivity in cobalt oxide Na$_{x}$CoO$_2$$ \cdot$1.3H$_2$O has attracted considerable attention \cite{Takada}. The new material consists of two-dimensional CoO$_2$ layers that bear similarity to high temperature (high-$T_c$) copper oxide superconductors. A simple but wide-spread view of the electronic state is that increasing the sodium content $x$ adds $x$ electrons to the 
spin 1/2 network in the CoO$_2$ layer \cite{Baskaran,WangLeeLee}, which is similar to the case of doping the spin 1/2 network in the CuO$_2$ layer of the high-$T_c$ cuprates. However, there is a large difference between these two classes of transitional-metal oxides. Namely, Co forms a triangular lattice in the cobaltate, instead of square lattice in the cuprates. Because of this crystal structure, exotic superconducting states are expected in the cobaltate. So far, chiral $d$-wave, $p$-wave,  $f$-wave and  $i$-wave  states \cite{Baskaran,WangLeeLee,TanakaHu,Ogata,Lee,Ikeda,Kuroki0,Kuroki}, as well as odd-frequency $s$-wave state \cite{Johanes} have been proposed theoretically. The $d$- and $i$-wave states are spin singlet states, while the remainder are spin triplet states.

Indeed, the nuclear spin-lattice relaxation rate ($1/T_1$) measurements by us indicate that the superconducting state is unconventional; $1/T_1$ is proportional to $T^3$ below the transition temperature $T_c$, suggesting line nodes in the gap function \cite{Fujimoto, Zheng}. However, the symmetry of the spin pairing has been unclear. Previous  measurements of the angle-averaged Knight shift in powder samples have resulted in   controversial conclusions \cite{Kato,Kobayashi}. Kato {\it et al} reported no change of the shift below $T_c$  \cite{Kato}, while Kobayashi {\it et al}  found a decrease of the shift in a powder sample \cite{Kobayashi}. The former suggested  triplet pairing, while the latter concluded to $s$-wave pairing. The difficulty in growing the samples,  which often accompany a  secondary phase that gives rise to a spurious NMR peak overlapping  the main peak in question,  complicates the data analysis in power samples. Therefore, measurements of the spin susceptibility via the Knight shift in single crystals are required. 

In particular, the Knight shift in the $c$-axis direction is essential to distinguish various pairing states proposed theoretically  \cite{Baskaran,WangLeeLee,TanakaHu,Ogata,Lee,Ikeda,Kuroki0,Kuroki,Johanes}. This is because even for triplet pairing, the spin susceptibility in the $a$-axis direction can be reduced below $T_c$ but that along the $c$-axis direction should remain unchanged across $T_c$ \cite{Yanase}. Kobayashi {\it et al} have attempted to align small single crystal platelets \cite{Kobayashi2}, but their samples contain multiple phases with almost equal amount of non-superconducting composition that contaminated the NMR signal observed. 
 So far,  no work has been reported to determine the absolute value of the $c$-axis shift due to the spin susceptibility which is important in discussing the superconducting state properties. 

In this paper, we report the first complete data set of the $^{59}$Co Knight shift  in a high quality {\it single crystal} with $T_c$=4.25 K. We find that the spin contribution to the Knight shift is significantly large and anisotropic. The spin Knight shift along both the $a$-axis and $c$-axis decrease below $T_c$. Our results indicate unambiguously that the spins that form the Cooper pairs  in the superconducting state are anti-paralleled.


Single crystal of Na$_{0.42}$CoO$_2$$\cdot$1.3H$_2$O used in this study was grown by the traveling solvent floating zone (TSFZ) method, as described in a previous publication \cite{Chen}.  AC susceptibility measurement using the in-situ NMR coil indicates that the $T_c$ at zero magnetic field is 4.25 K (data not shown). 
NMR experiments were performed using a home-built phase-coherent spectrometer.  The NMR spectra were taken by changing the external magnetic field ($H$) at a fixed RF frequency and recording the echo intensity step by step. 



Figure 1 shows a typical example of the NMR spectra for $H\parallel$c-axis and $H\parallel$a-axis. As can been seen in Fig. 1(a), a sharp central transition line accompanied by six satellites due to the nuclear quadrupole interaction, are observed. 
There is essentially no signal intensity between the sharp peaks, which indicates good quality of the single crystal.

\begin{figure}
\begin{center}
\includegraphics[scale=0.4]{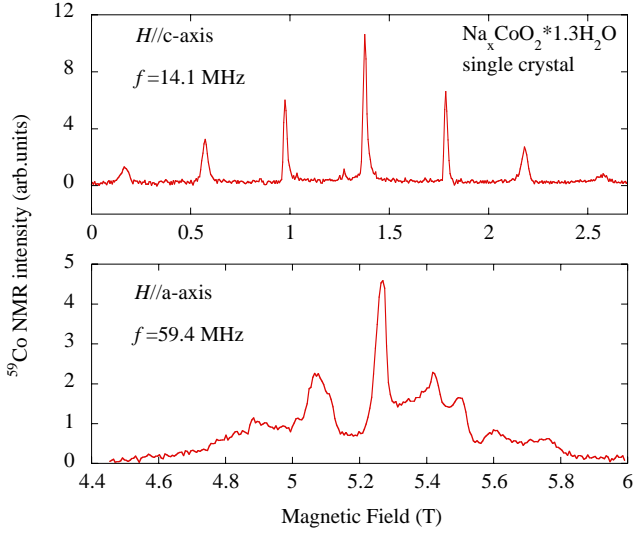}
\caption{(Color online) $^{59}$Co NMR spectra in Na$_{0.42}$CoO$_2$$\cdot$1.3H$_2$O at $T$=4.2 K with $H\parallel$c-axis at 14.1 MHz (upper panel) and with $H\parallel$a-axis at 59.4 MHz (lower panel).}
\label{fig:1}
\end{center}
\end{figure}

Figure 2 shows the temperature dependence of the Knight shift $K_a$ with the field applied along the $a$-axis, and $K_c$ with the field applied along the $c$-axis. 
$K_c$ was determined from the central peak as well as from the mid-point between the two first satellites; results obtained by the two procedures agree well \cite{note0}. $K_a$ was determined from the central transition, by taking into account the shift due to quadrupolar interaction effect \cite{Abragam}.
Upon decreasing the temperature from 260 K, both  $K_a$ and $K_c$ decrease, but tend to be saturated below $T$=100 K before superconductivity sets in. 
The decrease of the Knight shift with decreasing temperature is consistent with the bulk susceptibility results reported by Foo {\it et al} \cite{Foo} and also for the  sample used in the present study \cite{Alloul}, although a 
previous measurement  has failed to detect the temperature dependence \cite{Imai}. 
Below, we  show that the spin shift is quite large and can be used to distinguish the pairing symmetry in the superconducting state.

\begin{figure}
\begin{center}
\includegraphics[scale=0.4]{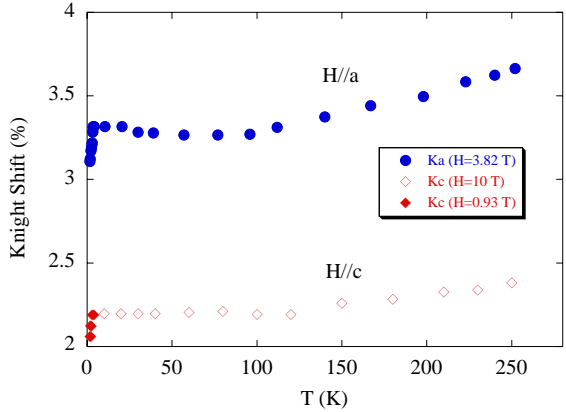}
\caption{(Color online) Temperature dependence of the Knight shift $K_a$ with the field applied along the $a$-axis,  and $K_c$ with the field applied along the $c$-axis.}
\label{fig:2}
\end{center}
\end{figure}

The Knight shift is composed of the spin part $K^s$ due to the spin susceptibility ($\chi^s$) and the orbital  part $K^{orb}$ due to the orbital susceptibility ($\chi^{orb}$) which is $T$-independent. 
\begin{eqnarray}
K = K^{orb}+K^s
\end{eqnarray}
\begin{eqnarray}
K^s=A^{hf}\chi^s
\end{eqnarray}
where $A^{hf}$ is the hyperfine coupling constant between the nuclear and the electron spins. Figure 3 shows the plot of $K_c$ versus $K_a$. The straight line is a fitting to the data which gives $K_c$=0.5$K_a$+const., which means that $K^s_{a}=2K^{s}_{c}$.
The temperature dependence of the Knight shift  is similar to the case of electron-doped cuprate superconductor (Pr$_{0.91}$La$_{0.09}$)$_2$CeCuO$_4$, in which $K_a$ also decreases with decreasing temperature and becomes a constant below $T$=60 K \cite{Zheng1}.

\begin{figure}
\begin{center}
\includegraphics[scale=0.4]{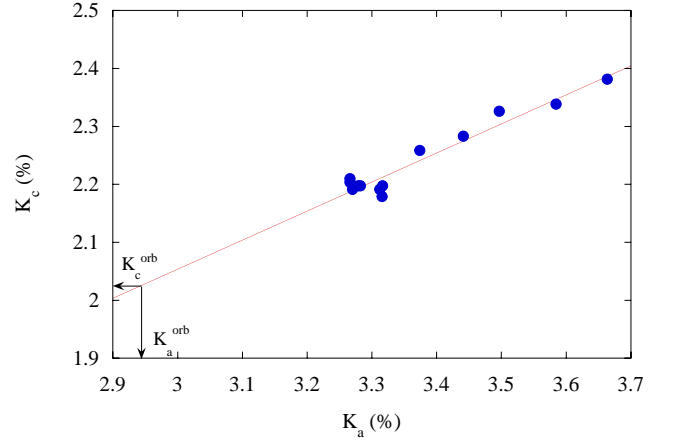}
\caption{(Color online) The plot of $K_c$ vs $K_a$. The straight line is the best fit to the data which gives $K^s_{a}=2K^{s}_{c}$. The arrows indicate the estimated orbital contributions to the shifts (see text).}
\label{fig:3}
\end{center}
\end{figure}

Let us focus  on the temperature dependence of the Knight shift below $T_c$. First, both $K_a$ and $K_c$ decrease below $T_c(H)$, as can be seen in
Figure 4 that shows the enlarged part of the shift below $T$=5 K for $H$(3.82 T)$\parallel a$-axis and $H$(0.93 T)$\parallel c$-axis. $T_c(H\parallel a$)  and $T_c(H\parallel c$) at which the shift decreases is 3.5 K and 3.1 K, respectively, which is in good agreement with the report by Chou {\it et al} \cite{Chou}. Figure 5 compares two of the spectra above and below $T_c$, respectively, for both field configurations.  
The reduction of $K_a$ and $K_c$ at $T$=1.6 K amounts to 0.2\% and 0.13\%, respectively. The smaller reduction of $K_c$ is mostly due to  the smaller spin part $K^s_c$. 
For $H\parallel c$-axis, we have also carried out measurements at 14.1 MHz ($H\sim$1.38 T), and found detectable shift change between $T_c (H)$ and $T$=1.6 K.  Namely, $\Delta K_c$=$K$ (2.8 K) - $K$ (1.6 K) = 0.05\%. We have further measured the shift at  7.1 MHz ($H\sim$0.69 T) in another sample with $T_c$=3.6 K,  and found $\Delta K_c$=$K$ (3.0 K) - $K$ (1.6 K) = 0.15\%. In all cases,
we have confirmed that the   $^{23}$Na NMR spectrum  has a full width at half maximum (FWHM) of 20 kHz, and does not move appreciablly across $T_c$ within the  experimental accuracy. Assuming that one can detect a change of one tenth  the FWHM in the $^{23}$Na spectrum, we estimate that the contribution to the Knight shift due to the diamagnetism effect ($K^{dia}$) in the vortex state is no larger than 0.02\%  \cite{note}. 
Therefore the reduction of  $K_a$ and $K_c$ below $T_c(H)$ is due predominantly to the reduction of the spin susceptibility. This result indicates that the spin pairing in the superconducting state is in the {\it singlet} state. 

\begin{figure}
\begin{center}
\includegraphics[scale=0.4]{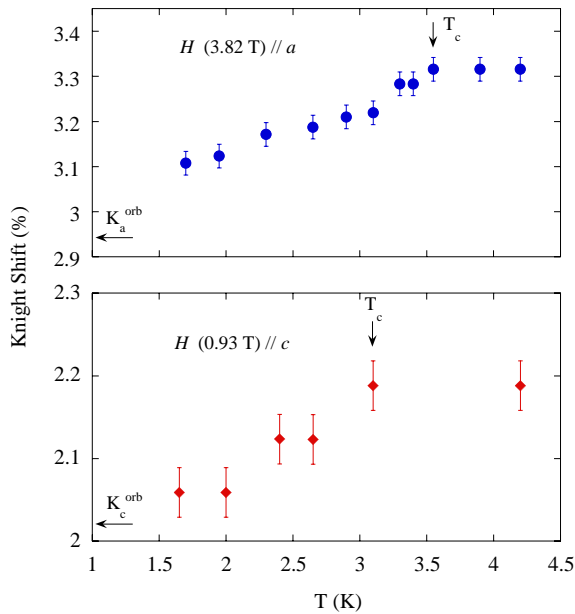}
\caption{(Color online) The enlarged part of the $^{59}$Co Knight shift at low temperatures. The vertical arrows indicate $T_c(H)$ at the respective external fields. The horizontal arrows indicated $K_{a,c}^{orb}$, the shift due to orbital susceptibility (see text for the estimation).}
\label{fig:4}
\end{center}
\end{figure}

\begin{figure}
\begin{center}
\includegraphics[scale=0.4]{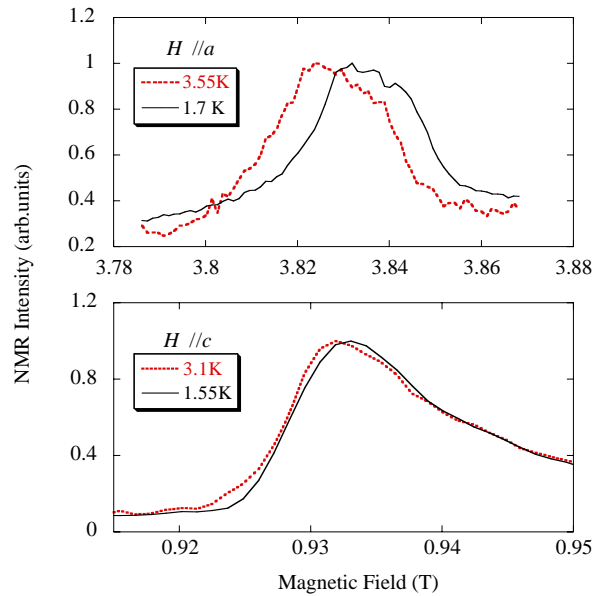}
\caption{(Color online) Comparison of the spectra above $T_c$ (broken curves) and below $T_c$ (solid curves) at different field configurations.}
\label{fig:5}
\end{center}
\end{figure}

Among the existing theoretical models, the $p$-wave \cite{TanakaHu}, $f$-wave \cite{Ogata,Lee,Ikeda,Kuroki0} states and the odd-frequency $s$-wave state \cite{Johanes} are spin triplet states, while the chiral $d$-wave state \cite{Baskaran,WangLeeLee} is a spin singlet but fully gapped state. They are not in  agreement with our finding, but  probably the $i$-wave state \cite{Kuroki} is an exception. However, a pure $d$-wave state with line nodes in the gap function appears to be consistent with both the Knight shift data and the $T^3$ dependence of $1/T_1$ \cite{Zheng}. Generally speaking, a state with lower angular momentum ({\it i.e.} $d$-wave) is energetically more favorable compared to its higher angular momentum counterparts ({\it i.e.} $i$-wave).  In addition, there are two conditions pertinent to the present case that favor a pure $d$-wave state. Firstly, although a fully-gapped $d_1+id_2$ state is predicted for an exact triangular lattice \cite{Baskaran,WangLeeLee},  
anisotropy (10$\sim$20\%) of the lattice will stabilize a pure $d$-wave state \cite{Watanabe}. Secondly, it has been found that the spin order in Na$_{0.5}$CoO$_2$ is antiferromagnetic, with alternative rows of Co$^{4+}$ (s=1/2) and Co$^{3+}$ (s=0) in the CoO$_2$ plane \cite{Neutron}. In other words, the triangles with regards to the spin configuration are effectively squared. In fact, both  the Knight shift found in this study and the temperature dependence of $1/T_1T$ in the normal state in Na$_{x}$CoO$_2$$\cdot$1.3H$_2$O ($x=0.25\sim0.35$) \cite{Zheng} point to the presence of antiferromagnetic correlations, which would favor $d$-wave pairing.

Next we attempt to separate the $K^{s}$ and $K^{orb}$ from the observed shift data, and argue that the data below $T_c$ can be consistently accounted for by an anisotropic superconducting state with  nodes in the gap function and with spin singlet pairing.
In $s$-wave superconductors with isotropic gap, the spin susceptibility outside the vortex cores vanishes completely at a temperature far below $T_c$. In anisotropic superconductors with nodes in the gap function, however, the spin susceptibility outside the vortex cores can remain a finite value since the quasiparticles state can be extended from the vortex core along the nodal directions \cite{Volovic}. Volovik pointed out that the density of such delocalized quasiparticle state, $N_V$, is proportional to $\sqrt{H/H_{c2}}$ where $H_{c2}$ is the upper critical field  \cite{Volovic}. Indeed, in an overdoped cuprate superconductor, it has been found that  $N_V/N_0\sim0.7\sqrt{H/H_{c2}}$ (ref.\cite{Zheng2}), where $N_0$ is the density of states (DOS) in the normal state. Note that the coefficient 0.7 is less than 1, because NMR only probes the DOS outside the vortex cores. If we use the data of $H_{c2}^a$(0)=8 T and $H_{c2}^c$(0)=4 T by Chou {\it et al} \cite{Chou} and the same $N_V$  as in the cuprate, we obtain $K^{orb}_a$ =2.94\% from the relation $0.7\sqrt{H/H_{c2}}(K_a$(4.2 K)-$K^{orb}_a )=K_a$(1.6 K)-$K^{orb}_a$. Once $K^{orb}_a$ is known, one obtains readily $K^{orb}_c$ =2.02\% from the $K_c$ vs $K_a$ plot in Fig. 3. Such estimate  consistently explains the $K_c$ result; the $H$-induced shift at the lowest temperature due to $N_V(H\parallel c)$ is calculated to be 0.05\%, which is in fair agreement with the experimental result of 0.04$\pm$0.03\%. Note that  $K_a^s\geq$0.72\%, which is about two times the spin shift in cuprate superconductors \cite{Zheng1,Slichter}
Finally, we comment that the estimated $K_{a,c}^{orb}$ agrees fairly well with that independently  estimated from the  Knight shift vs DC susceptibility ($K vs \chi$) plot \cite{MatanoZheng}.



In conclusion,  the $^{59}$Co NMR measurements in a single crystal  of Na$_{x}$CoO$_2$$\cdot$1.3H$_2$O ($T_c$=4.25 K) allowed us, for the first time, to obtain a complete data set of the Knight shift. We found that the shift due to the spin susceptibility is substantially large (0.72\% along the $a$-axis) and anisotropic, with $K^s_a=2K^s_c$. Both $K^s_a$ and $K^s_c$ decrease below $T_c$,  indicating that the spin pairing of the Cooper pairs is in the singlet state. Our work that  settled  the electron pairing symmetry in the superconducting state should help pave the first step toward understanding the mechanism of the superconductivity  in  this new class of materials.
 

We thank M. Nishiyama for assistance in some of the measurements. This work was  partially supported by  MEXT grants for scientific research.


\begin{thebibliography}{}

\bibitem{Takada}
K. Takada, H. Sakurai, E. Takayama-Muromachi, F. Izumi, R.A. Dilanian and T. Sasaki, Nature {\bf 422}, 53 (2003).
%
\bibitem{Baskaran}
G. Baskaran, Phys. Rev. Lett. {\bf 91}, 097003 (2003).
%
\bibitem{WangLeeLee}
Q.H. Wang, D.H. Lee and P.A. Lee, Phys. Rev. {\bf B 69}, 092504 (2004).
%
\bibitem{TanakaHu}
A. Tanaka and X. Hu,   Phys. Rev. Lett. {\bf 91}, 257006 (2003).
%
\bibitem{Ogata}
Y. Tanaka, Y. Yanase, and M. Ogata, J. Phys. Soc. Jpn. {\bf 73}, 319 (2004).
%
\bibitem{Lee}
O. I. Motrunich  and P.A. Lee, Phys. Rev. 70, 024514 (2004).
%
\bibitem{Ikeda}
H. Ikeda, Y. Nishikawa, K. Yamada, J. Phys. Soc. Jap.
{\bf 73},  17 82004).
%
\bibitem{Kuroki0}
K. Kuroki, Y. Tanaka, and R. Arita,
Phys. Rev. Lett. {\bf 93}, 077001 (2004)
%
\bibitem{Kuroki}
K. Kuroki, Y. Tanaka and R. Arita, Phys. Rev. {\bf B 71},  024506 (2005).


%
\bibitem{Johanes}
M. D. Johannes, I. I. Mazin, D. J. Singh, and D. A. Papaconstantopoulos, Phys. Rev. Lett. {\bf 93}, 097005 (2004).
%
\bibitem{Fujimoto}
T. Fujimoto, G. - q. Zheng, Y. Kitaoka, R.L. Meng, J. Cmaidalka, and C.W. Chu, Phys. Rev. Lett. {\bf 92}, 047004 (2004).
%
\bibitem{Zheng}
G. - q. Zheng, K. Matano, R.L. Meng, J. Cmaidalka, and C.W. Chu, J. Phys.: Condens. Matter {\bf 18}, L63 (2006).
%
\bibitem{Kato}
M. Kato, C. Michioka, T. Waki, Y. Itoh, K. Yoshimura, K. Ishida, H. Sakurai, E Takayama-Muromachi, K. Takada and T. Sasaki, J. Phys.: Condens. Matter, {\bf 18}, (2006) 669.

%
\bibitem{Kobayashi}
Y. Kobayashi, M. Yokoi and M. Sato, J. Phys. Soc. Jpn. {\bf 72}, 2453 (2003).
%

\bibitem{Yanase}
Y. Yanase, M. Mochizuki and M. Ogata, J. Phys. Soc.  Jap. {\bf 74},   2568 (2005). 

\bibitem{Kobayashi2}
Y. Kobayashi, H. Watanabe, M. Yokoi, T. Moyoshi, Y. Mori and M. Sato, J. Phys. Soc. Jpn. {\bf 74}, 1800 (2005).

%
\bibitem{Chen}
D.P. Chen, H. C. Chen, A. Maljuk, A. Kulakov, H. Zhang, P. Lemmens, and C. T. Lin,  Phys. Rev. {\bf B 70}, 024506 (2004)
%
\bibitem{note0}
The nuclear quadrupole resonance frequency, $\nu_Q$, is temperature independent below $T$=90 K. 

\bibitem{Abragam}
A. Abragam, {\it Principles of Nuclear Magnetism}, Oxford University Press (1961, Oxford).

%

\bibitem{Foo}
M.L. Foo, Y. Wang, S. Watauchi, H. W. Zandbergen, T. He, R. J. Cava, and N. P. Ong, Phys. Rev. Lett. {\bf 92}, 247001 (2004).

\bibitem{Alloul}
D.P. Chen and C.T. Lin, unpublished.
%
\bibitem{Imai}
F. L. Ning, T. Imai, B. W. Statt, and F. C. Chou, 
Phys. Rev. Lett. {\bf 93}, 237201 (2004).

%
\bibitem{Zheng1}
G. - q. Zheng, T. Sato, Y. Kitaoka, M. Fujita and K. Yamada, Phys. Rev. Lett. {\bf 90}, 197005 (2003).
%
%
\bibitem{Chou}
F.C. Chou, J. H. Cho,  P. A. Lee,  E. T. Abel,  K. Matan,  and Y. S. Lee, Phys. Rev. Lett. {\bf 92}, 157004 (2004).
%
\bibitem{note}
From the central peak that consists almost equally of the contribution from superconducting and non-superconducting compositions, Kobayashi {\it et al} have reported a shift change of 0.04$\pm$0.01\% between $T$=3 K and 1.5 K \cite{Kobayashi2}, which is less than one quarter (1/4) the value we found. Note that such shift change of 0.04$\pm$0.01\% is comparable to the estimated $K^{dia}$. In fact, very recently Ihara {\it et al} (J. Phys. Soc. Jap. {\bf 75}, (2006) 013708) claimed that $K_c^s$ does not decrease  beyond such $K^{dia}$. 

%
\bibitem{Watanabe}
Watanabe, H. Yokoyama, Y. Tanaka, J. Inoue, M. Ogata, Cond-mat/0411711
%
\bibitem{Neutron}
G. Gasparovic, R. A. Ott, J.-H. Cho, F. C. Chou, Y. Chu, J. W. Lynn, Y. S. Lee,  cond-mat/0508158.
%
\bibitem{Volovic}
G.E. Volovik, JETP Letters {\bf 58}, 469 (1993).
%
\bibitem{Zheng2}
G. - q. Zheng, H. Ozaki,  Y. Kitaoka,  P. Kuhns,  A. P. Reyes,  and W. G. Moulton, Phys. Rev. Lett. {\bf 88}, 077003 (2002).
%
\bibitem{Slichter}
S. E. Barrett, D. J. Durand, C. H. Pennington, C. P. Slichter, T. A. Friedmann, J. P. Rice, and D. M. Ginsberg, 
Phys. Rev. {\bf B 41}, 6283 (1990).

\bibitem{MatanoZheng}
K. Matano and G. - q. Zheng, to be published.

\end{thebibliography}
\end{document}